\documentstyle[psfig]{mn}

\title[Radially truncated galactic discs]{Radially truncated galactic
discs\thanks{Based on observations obtained at the European Southern
Observatory, La Silla, Chile}}

\author[Richard de Grijs, Michiel Kregel and Karen H.  Wesson]{Richard
de Grijs$^{1,2,}$\thanks{E-mail: grijs@ast.cam.ac.uk}, Michiel
Kregel$^3$ and Karen H.  Wesson$^{1,}$\thanks{Present address: Center
for Hydrologic Science, Duke University, 106 Old Chemistry, Box 90230,
Durham, NC 27708, USA}
\\ 
$^1$ Astronomy Department, University of Virginia, PO Box 3818,
Charlottesville, VA 22903, USA \\ 
$^2$ Institute of Astronomy, University of Cambridge, Madingley Road,
Cambridge CB3 0HA \\
$^3$ Kapteyn Astronomical Institute, University of Groningen, PO Box
800, 9700 AV Groningen, the Netherlands
}

\date{Received date; accepted date}
\pubyear{2000}

\begin{document}
\maketitle

\begin{abstract}
We present the first results of a systematic analysis of radially
truncated exponential discs for four galaxies of a complete sample of
disc-dominated edge-on spiral galaxies.  \\
The discs of our sample galaxies are truncated at similar radii on
either side of their centres.  With possible the exception of the disc
of ESO 416-G25, it appears that the truncations in our sample galaxies
are closely symmetric, in terms of both their sharpness and the
truncation length.  However, the truncations occur over a larger region
and not as abruptly as found in previous studies.  \\
We show that the truncated luminosity distributions of our sample
galaxies, if also present in the mass distributions, comfortably meet
the requirements for longevity.  The formation and maintenance of disc
truncations are likely closely related to stability requirements for
galactic discs. 
\end{abstract}

\begin{keywords}
galaxies: formation -- galaxies: fundamental parameters -- galaxies:
photometry -- galaxies: spiral -- galaxies: structure 
\end{keywords}

\section{Introduction}
\label{exptrunc.sec}

It is well-known from Freeman's (1970) work that the radial light
distribution of the stellar component of high surface brightness
galactic discs can be approximated by an exponential law of the form
\begin{equation}
\label{freeman.eq}
L(R) = L_0 \exp (- R / h_R)
\end{equation}
where $L_0$ is the luminosity density in the galactic centre, {\it R}
the galactocentric distance and $h_R$ the disc scalelength. 

However, for a few prominent high surface-brightness edge-on galaxies,
it has initially been shown by van der Kruit \& Searle (1981a,b,
1982a,b, hereinafter KS1--4), that at some radius $R_{\rm max}$ (the
truncation or cut-off radius of the galactic disc) the stellar
luminosity distribution disappears asymptotically into the background
noise (see also Jensen \& Thuan 1982, Sasaki 1987, Morrison, Boroson \&
Harding 1994, Bottema 1995, Lequeux, Dantel-Fort \& Fort 1995, N\"aslund
\& J\"ors\"ater 1997, Fry et al.  1999, Pohlen et al.  2000a,b).  In
fact, the truncation of galactic discs does not occur instantly but over
a small region, where the luminosity decrease becomes much steeper,
having exponential scalelengths of order or less than a kiloparsec,
opposed to several kpc in the exponential disc part (e.g., KS1--4,
Jensen \& Thuan 1982, Sasaki 1987, Abe et al.  1999, Fry et al.  1999). 

An independent approach to obtain the statistics of truncated galactic
discs, using a sample of galaxies selected in a uniform way, is needed
in order to better understand the overall properties and physical
implications of this feature.  In this paper we present the first
results of a systematic analysis of disc truncations for a pilot sample
of four ``normal'' spiral galaxies, drawn from the statistically
complete sample of edge-on disc-dominated galaxies of de Grijs (1998). 
In Kregel, van der Kruit \& de Grijs (2001) we will present a
re-analysis of the global disc structures of the entire de Grijs (1998)
sample, including a systematic analysis of the occurrence of truncated
galactic discs. 

Edge-on galaxies are particularly useful for the study of truncated
galactic discs: since these disc cut-offs usually occur at very low
surface brightness levels, they are more readily detected in highly
inclined galaxies, where we can follow the light distributions out to
larger radii.  In Sect.  \ref{technical.sect} we outline the sample
selection and our approach and methodology. 

A detailed error discussion for our pilot sample is given in Sect. 
\ref{results.sect}.  The most important science driver for the study of
truncated galactic discs, discussed in Sect.  \ref{dynamics.sect}, is
that if the truncations seen in the stellar light are also present in
the mass distribution, they would have important dynamical consequences
at the disc's outer edges. 

\section{Global Approach}
\label{technical.sect}

\subsection{Pilot sample}

We selected four galaxies from the statistically complete sample of
disc-dominated edge-on galaxies of de Grijs (1998) for this pilot study. 

The galaxies in the parent sample were selected to:
\begin{itemize}
\item have inclinations $i \ge 87^\circ$;
\item have blue angular diameters $D_{25}^B \ge 2.'2$; and
\item be non-interacting and undisturbed S0 -- Sd galaxies.
\end{itemize}

We required these pilot galaxies to have relatively high signal-to-noise
(S/N) ratios out to large galactocentric distances, thus allowing us to
better determine the occurrence of a possible truncated disc at large
radii.  Of the parent sample of 48 edge-on disc galaxies, $\sim 25$ met
our overall high-S/N selection criteria in {\em all} of the {\it B, V}
and {\it I} observations.  The four galaxies selected for our pilot
sample were chosen randomly from among the larger disc galaxies because
of their small bulge-to-disc ratio and well-defined, regular disc
component, which was only negligibly or minimally affected by a central
dust lane (cf.  Fig.  2 in Chapter 9 of de Grijs 1997). 

\begin{figure*}
\vspace*{1.9cm}
\psfig{figure=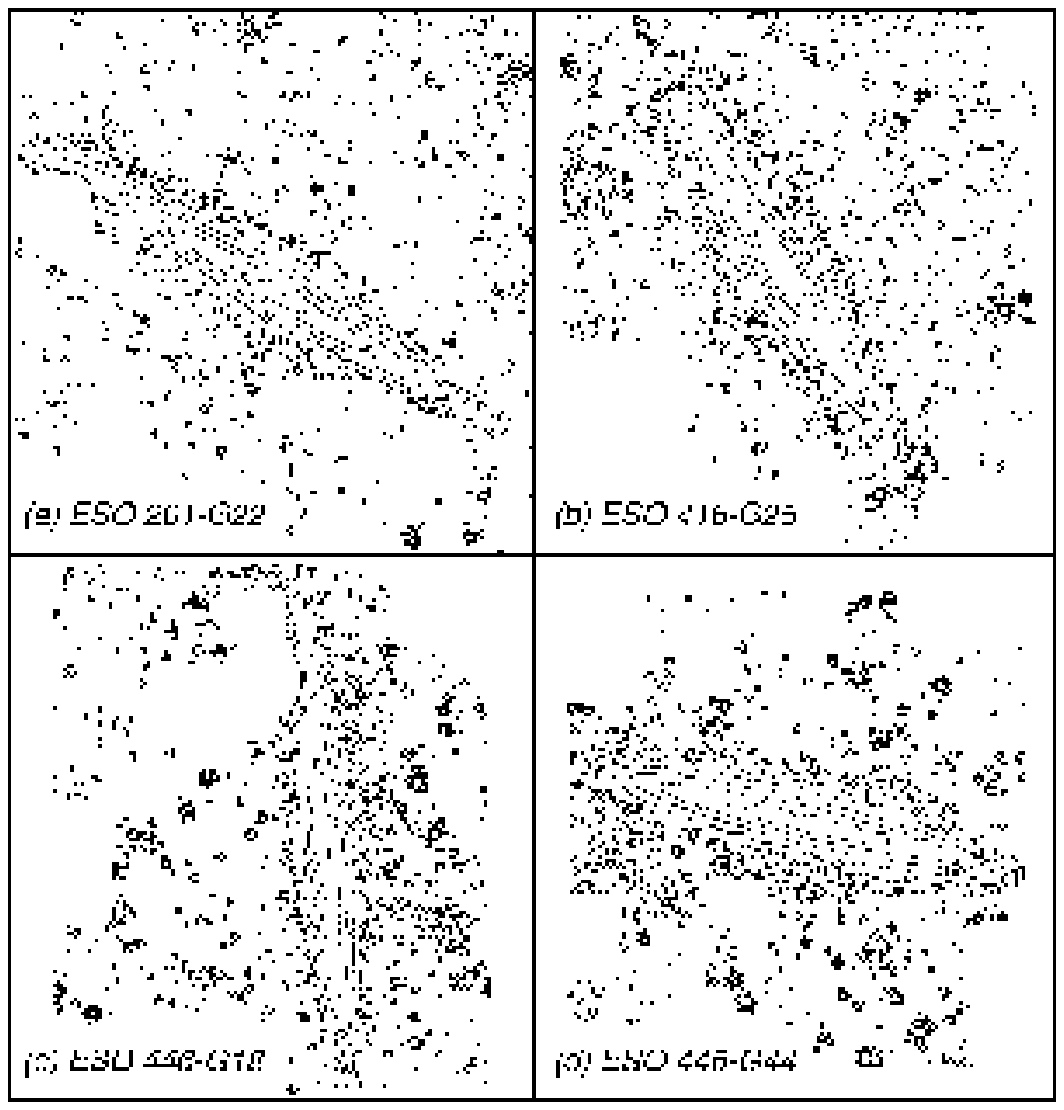}
\vspace*{0.8cm}
\caption[]{\label{contours.fig}Optical {\it I}-band isophotes of the
four edge-on disc galaxies in our pilot sample.  North is up, East to
the left; each panel is approx.  $3'$ on a side.  The contours are
spaced by 1.0 mag, starting from the lowest contours at 23.5 {\it I}-mag
arcsec$^{-2}$.  For reasons of clarity, the brightest foreground stars
(that may be potentially contaminating the galactic discs) have been
masked out.  Note that an indication of the presence of disc truncations
is already given by the rounded contours at the discs' outer edges.}
\end{figure*}

The basic physical properties of these pilot sample galaxies are
summarised in Table \ref{pilot.tab}.  The observational properties were
taken from de Grijs (1997, 1998).  The derived properties are obtained
in Sect.  \ref{indiv.sect}.  They are based on the detailed modelling of
the galactic luminosity density distributions projected on the plane of
the sky using the method described in Kregel et al.  (2001).  The
detailed observational logs of and data reduction techniques applied to
our {\it B, V} and {\it I}-band observations are summarised in de Grijs
(1998).  Fig.  \ref{contours.fig} displays the {\it I}-band contours of
these galaxies. 

{
\begin{table*}

\caption[]{\label{pilot.tab}{\bf Basic properties of the pilot sample
galaxies}\\

Columns: (1) Galaxy name (Lauberts \& Valentijn 1989; ESO-LV); (2) and
(3) Coordinates; (4) Revised Hubble Type; (5) Blue major axis diameter,
$D_{25}^B$; (6) Passband observed in; (7)
apparent magnitude, corrected for foreground
extinction; (8) extrapolated edge-on disc central surface
brightness; (9) Exponential scaleheight; (10) Exponential scalelength.}

\begin{center}
\tabcolsep=1mm

\begin{tabular}{cccccccccc}
\hline
Galaxy & \multicolumn{2}{c}{RA (J2000) Dec} & Type & $D_{25}^B$ &
Passband & $m_0$ & $\mu_0$ & $h_z$ & $h_R$ \\
\noalign{\vspace{2pt}}
\cline{2-3}\cline{9-10}
\noalign{\vspace{1pt}}
 & $(^{\rm h \; m \; s})$ & $(^{\circ}$ $'$ $'')$ & (T) & $(')$ & &
(mag) & (mag arcsec$^{-2}$) & \multicolumn{2}{c}{$('')$} \\
(1) & (2) & (3) & (4) & (5) & (6) & (7) & (8) & (9) & (10) \\
ESO 201-G22 & 04 08 59.3 & $-$48 43 42 & 5.0 & 2.52 & $B$ & $14.06 \pm 0.07$ & $20.54 \pm 0.07$ & $2.4 \pm 0.1$ & $26.9 \pm 1.5$ \\
            &            &             &     &      & $V$ & $              $ & $20.39 \pm 0.10$ & $2.4 \pm 0.1$ & $25.2 \pm 1.1$ \\
            &            &             &     &      & $I$ & $13.10 \pm 0.03$ & $19.23 \pm 0.08$ & $2.6 \pm 0.2$ & $23.3 \pm 1.3$ \\
ESO 416-G25 & 02 48 40.8 & $-$31 32 07 & 3.0 & 2.35 & $B$ & $14.42 \pm 0.11$ & $21.41 \pm 0.03$ & $3.1 \pm 0.1$ & $29.2 \pm 1.4$ \\
            &            &             &     &      & $V$ & $              $ & $20.90 \pm 0.08$ & $3.5 \pm 0.2$ & $25.8 \pm 1.3$ \\
            &            &             &     &      & $I$ & $12.52 \pm 0.03$ & $19.87 \pm 0.05$ & $3.6 \pm 0.2$ & $22.2 \pm 2.0$ \\
ESO 446-G18 & 14 08 37.9 & $-$29 34 20 & 3.0 & 2.52 & $B$ & $15.12 \pm 0.05$ & $20.72 \pm 0.10$ & $2.2 \pm 0.1$ & $22.5 \pm 1.3$ \\
            &            &             &     &      & $V$ & $              $ & $20.03 \pm 0.08$ & $2.2 \pm 0.1$ & $19.5 \pm 0.9$ \\
            &            &             &     &      & $I$ & $12.71 \pm 0.02$ & $18.49 \pm 0.10$ & $2.0 \pm 0.2$ & $18.1 \pm 1.4$ \\
ESO 446-G44 & 14 17 49.3 & $-$31 20 55 & 6.0 & 2.67 & $B$ & $14.83 \pm 0.03$ & $20.39 \pm 0.10$ & $2.1 \pm 0.1$ & $25.1 \pm 0.7$ \\
(= IC 4393) &            &             &     &      & $V$ & $              $ & $19.76 \pm 0.07$ & $2.0 \pm 0.1$ & $24.5 \pm 0.6$ \\
            &            &             &     &      & $I$ & $12.51 \pm 0.04$ & $18.90 \pm 0.15$ & $2.7 \pm 0.2$ & $29.5 \pm 1.6$ \\
\hline
\end{tabular}
\end{center}
\end{table*}
}

\subsection{The adopted three-dimensional model}
\label{3Dmodel.sect}

We will approximate the three-dimensional (3D) luminosity density of the
discs of ``ideal'' spiral galaxies as a combination of independent
exponential light distributions in both the radial and the vertical
directions (see, e.g., de Grijs, Peletier \& van der Kruit [1997] for a
statistical approach to determine the latter behaviour), for all radii
excluding the region of truncation.  In view of the finite extent of the
cut-off region, $\delta$, and to avoid discontinuities in the luminosity
and density distributions, we will adopt a slightly modified version of
Casertano's (1983) mathematically convenient description for a ``soft
cut-off'' of the radial density distribution in the truncation region. 
His model assumes that in the region beyond ($R_{\rm max} - \delta$),
the radial luminosity density decreases linearly to zero:

\[
L(R) = L_0 \left\{ \begin{array}{ll}

\exp(-R / h_R), \\

\qquad\qquad\qquad \mbox{if } R \le (R_{\rm max} - \delta) \\

\exp\Bigl( -\Bigl( {{R_{\rm max} - \delta}  \over h_R} \Bigr) \Bigl[ {{1 - \Bigl( R -
(R_{\rm max} - \delta \bigr) \Bigr)} \over \delta} \Bigr], \\

\qquad\qquad\qquad \mbox{if } (R_{\rm max} - \delta) \le R \le R_{\rm max} \\

0, \qquad\qquad\quad \mbox{if } R > R_{\rm max}. \qquad\qquad\qquad\quad (2)
\end{array}
\right. \]

A model radial profile with a ``soft cut-off'' will therefore show an
exponentially declining disc component, displaying an abrupt steepening
at the onset of the truncation region at $(R_{\rm max} - \delta)$,
decreasing linearly until the background noise level is reached.  We
will use a slightly modified version of Eq.  (2), in which the radial
luminosity density decreases exponentially instead of linearly until it
disappears into the background noise.  Examples of our modified fitting
function are shown in Fig.  \ref{Iband.fig} for the current sample. 

The exact functional form of the radial luminosity density in the
cut-off region is unknown because of the limited spatial resolution and
low S/N ratio at these large galactocentric distances.  Thus, within the
observational uncertainties, Casertano's (1983) description of the
``soft cut-off'' does not deviate significantly from this simple
exponential form, also adopted by Jensen \& Thuan (1982) and N\"aslund
\& J\"ors\"ater (1997) to fit the truncated light profiles of NGC 4565. 
Furthermore, our approach, using the exponentially decreasing radial
functionality in the truncation region, will provide us with an
additional constraint on the shape of the truncation region compared to
Casertano's (1983), namely the scale length in the truncation region,
$h_{R,\delta}$. 

In the case of an edge-on orientation, the projection onto the plane of
the sky of our model radial exponential luminosity density distribution
is, for $R \le (R_{\rm max} - \delta)$, closely approximated by (KS1):
\addtocounter{equation}{+1}
\begin{equation}
\label{bessel.eq}
L(R) = L_0 {R \over h_R} K_1\Bigl({R \over h_R}\Bigr),
\end{equation}
where $K_1$ is the modified Bessel function of the first order.

The total projected 3D luminosity density is now given by
\begin{equation}
\label{2dmodel.eq}
L(R,z) = L(R) \exp(-z/h_z) \quad ,
\end{equation}
where {\it z} is the (vertical) distance from the galactic plane and
$h_z$ the exponential scaleheight.

\subsubsection{Differences with respect to earlier work}
\label{diffs.sect}

Except for a few detailed studies of individual large edge-on galaxies
(e.g., KS1, Jensen \& Thuan 1981, Sasaki 1987, N\"aslund \& J\"ors\"ater
1997, Abe et al.  1999, Fry et al.  1999), most previous analyses aimed
at determining disc truncations for statistically more meaningful
samples (e.g., KS3, Barteldrees \& Dettmar 1994, Pohlen et al.  2000a,b)
have adopted a number of {\it a priori} assumptions that may not be
fully justified.  In particular, these studies assumed that:

\begin{enumerate}
\item galactic discs are truncated at equal radii on either side of the
galactic centre, and
\item the radial surface brightness distributions disappear
asymptotically (``vertically'') into the background noise at $R_{\rm
max}$. 
\end{enumerate}

While the first assumption may approximate the observational situation
relatively closely (cf.  Sect.  \ref{asymm.sect} for the current pilot
sample), the surface brightness generally does not disappear
asymptotically into the noise for most of the well-resolved galaxies
studied to date. 

Forcing a fitting routine to satisfy both of these assumptions in
galaxies with slightly different truncation radii on either side of the
centre will often overpredict the surface brightness in one of the
truncation regions significantly (depending on the extent of this
region), even if the full projected 3D surface brightness distribution
is used.  This is clearly illustrated in Pohlen et al.  (2000b), in
particular in their {\it i}-band fits to the brightest profiles parallel
to the major axis of the galaxies IC2207, IC4393, ESO446-G18, ESO466-G01
and ESO578-G25.  Alternatively, if the truncation radii on either side
of the centre are similar but the truncation scalelength is relatively
long, forcing a 3D fitting routine to adopt sharp cut-offs as in the
latter assumption, will {\it under\ }predict the actual radius where the
galactic luminosity density disappears into the noise. 

In view of these considerations, we will avoid such {\it a priori}
assumptions in our approach; we will determine the actual truncation
radii independently on either side of the galactic centre, and
approximate the luminosity density distribution in the truncation region
by an exponentially decreasing function of radius. 

Examples of the fitting method for our pilot sample are presented in
Sect.  \ref{properties.sect}. 

\subsection{Surface brightness modelling}
\label{2D.sect}

Although the determination of the actual truncation radius $R_{\rm max}$
is relatively model-independent, for the detailed modelling of the
truncation region $(R_{\rm max} - \delta)$, it is crucially important to
determine accurate scale parameters (in particular scale lengths) for
the main, exponential disc component, cf.  Eqs.  (2) and (3). 

The surface brightness distributions of spiral galaxies often show
significant local deviations from the assumed smooth, large-scale model
distribution (3) (e.g., Seiden, Schulman \& Elmegreen 1984, Shaw \&
Gilmore 1990, de Jong 1995, and references therein).  This makes the
global applicability of radial disc scalelengths obtained from radial
profiles parallel to the major axes of edge-on galaxies very uncertain
(e.g., Knapen \& van der Kruit 1991, Giovanelli et al.  1994). 

Therefore, we will model the global disc structures using the full
projected luminosity density distributions of the galactic discs in our
sample, using Eqs.  (3) and (4), in the linear regime.  An elaborate 
description of the method, as well as extensive tests on artificial
images, will be presented in Kregel et al.  (2001).  This paper will
only address the occurrence of disc truncations in the de Grijs (1998)
sample, without embarking on a detailed analysis of their shapes,
however. 

A number of potential problems are foreseen regarding the applicability
of our simple model, Eq.  (4), to the observed luminosity distributions
of edge-on galaxies. 

First, it does not include a description of the truncation region.  A
physically motivated, or even an empirical description of this region
is, at this point, too premature and therefore this region will be
masked out before least-squares minimization. 

Secondly, the regions near the galactic planes are affected by
extinction.  Whilst these effects can be included by using a 3D
radiative transfer code (e.g., Kylafis \& Bahcall 1987, Xilouris et al. 
1997), we choose to mask the data in the regions near the planes
instead.

Thirdly, by adopting our simple model, Eq.  (4), we do not include the
effects of a truncated galactic disc along the line of sight.  However,
the effects of neglecting a line-of-sight truncation are small (see
Kregel et al.  2001): while it may cause us to underestimate the disc
scalelength of ESO446-G44 by $\sim 15$\%, the effect is $\le 8$\% for
the other galaxies in our sample, and is in practice counteracted by
residual extinction by dust, if any, at the {\it z}-heights used in our
study.  In addition, this effect is negligible for the accurate
determination of the disc {\it truncations} as such, the main aim of our
study. 

Before applying our fitting routine to the full observed luminosity
density distributions, the sky-subtracted images needed to be prepared. 
First, foreground stars and background galaxies were masked out. 
Secondly, profiles parallel to the minor axis were taken at various
distances along the major axis.  These were subsequently inspected for
extinction effects, leading to the masking out of the regions described
below for the individual galaxies. 

\subsubsection{Fits to the individual galaxies}
\label{indiv.sect}

We will now discuss the fits to the luminosity density of each galaxy
individually, thereby addressing a number of problems encountered in
each case.  In all fits the galactic centres were fixed at the values
determined by fitting ellipses to the {\it I}-band isophotes, using a
custom-written IRAF\footnote{The Image Reduction and Analysis Facility
(IRAF) is distributed by the National Optical Astronomy Observatories,
which is operated by the Association of Universities for Research in
Astronomy, Inc., under cooperative agreement with the National Science
Foundation.} package for galactic surface photometry (``{\sc
galphot}''). 

In each case, conservative estimates of the onset of the truncation
region and of the region near the galactic plane most affected by
extinction were made based on detailed visual examinations of the radial
and vertical luminosity distributions, respectively.  The inner
boundaries used for the radial fitting were chosen to minimize possible
bulge effects and will be discussed individually below.  Table
\ref{boundaries.tab} summarizes the radial ranges adopted for the disc
fits, as well as the vertical ranges excluded to avoid extinction
effects. 

\begin{figure}
\psfig{figure=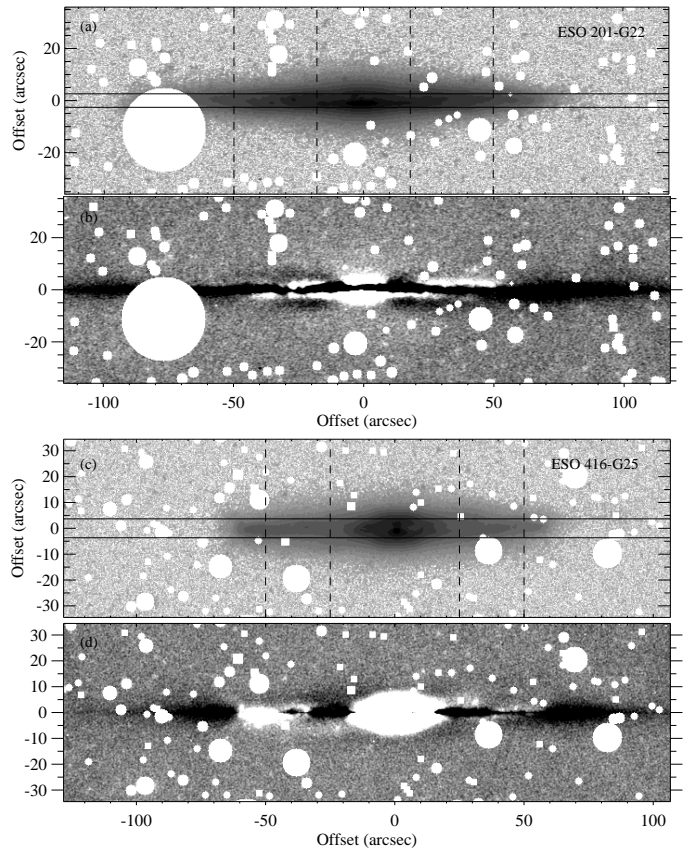,width=9.0cm}
\caption[]{\label{2D.fig}{\it (a)} -- Negative {\it I}-band image of ESO
201-G22, after removal of foreground stars and background galaxies.  The
radial fitting boundaries are indicated by the dashed lines; the dust
mask covering the region close to the plane affected by extinction is
bracketed by the solid lines; {\it (b)} -- Residuals for ESO 201-G22
after subtracting the model, greyscale levels range from $-6
\sigma_{\rm background}$ (black) to $+6 \sigma_{\rm background}$
(white) {\it (c)} and {\it (d)} -- ESO 416-G25, as (a) and (b)}
\end{figure}

Fig.  \ref{2D.fig} shows the {\it I}-band images and residual emission
(disc model subtracted from the observations) for ESO 201-G22 and ESO
416-G25.  Table \ref{pilot.tab} contains the resulting global scale
parameters.  The associated uncertainties are the observational errors,
estimated by comparing results from several similar fits in which we
adjusted the boundaries of the radial fitting range by 10--20\%; the
formal errors were in general less than 1\%. 

{
\begin{table}

\caption[]{\label{boundaries.tab}{\bf Fitting regions}\\ 
Columns: (1) Galaxy name; (2) and (3) Radial fitting range (arcsec); (4)
and (5) Vertical region around the galactic plane excluded from the fits
$(h_z)$.}

\begin{center}
\tabcolsep=1mm

\begin{tabular}{cccccc}
\hline
Galaxy & \multicolumn{2}{c}{Radial fitting} & &
\multicolumn{2}{c}{Galactic plane} \\
\cline{2-3}\cline{5-6}
\noalign{\vspace{2pt}}
& $r_{\rm min}$ & $r_{\rm max}$ & & $z_{\rm min}$ & $z_{\rm max}$ \\
(1) & (2) & (3) & & (4) & (5) \\
ESO 201-G22 & 15 & 50 & & --1.0 (S) & 1.0 (N) \\
ESO 416-G25 & 25 & 50 & & --1.0 (E) & 1.0 (W) \\
ESO 446-G18 & 18 & 55 & & --2.0 (E) & 0.5 (W) \\
ESO 446-G44 & ~~0 & 36 && --1.5 (S) & 1.5 (N) \\
\hline
\end{tabular}
\end{center}
\end{table}
}

\noindent
{\it ESO 201-G22} -- Figs.  \ref{2D.fig}a and b clearly show the bulge
component and, just outside the masked region, the effects of either
residual extinction near the galactic plane or, more likely, an
additional disc component.  The residuals in the fitted region do not
show large systematic effects; the bulge contribution to the
disc-dominated fitting range is negligible.  The negative residuals
extending to the edges of these figures clearly show the presence of a
truncation in the galactic light distribution. 

\noindent 
{\it ESO 416-G25} -- This is the earliest-type galaxy in our pilot
sample.  The residuals after subtracting the disc-only fit (Fig. 
\ref{2D.fig}d) do not appear to be systematic in the region where the
fit was done, and their amplitude is small (r.m.s.  residual $\approx
1.7\ \sigma_{\rm background}$).  Although we included a bulge component,
the 6-parameter fit proved to be very unstable. 

\noindent
{\it ESO 446-G18} -- Figs.  \ref{contours.fig}c and \ref{bgcheck.fig}
show that this galaxy is not exactly edge-on.  Extinction predominantly
affects the eastern side.  A dust mask placed symmetrically with respect
to the major axis is therefore not appropriate.  The residuals are
relatively large (r.m.s.  residual $\approx 3.3\ \sigma_{\rm
background}$) but do not appear to be systematic in the region where the
fit was done.  A fit including an exponential bulge did not converge. 

\noindent
{\it ESO 446-G44} -- Surface brightness profiles of ESO 446-G44 do not
reveal any bulge component.  They do show, however, that ESO 446-G44 may
not be exactly edge-on.  Again, the residuals in the region where we
applied our fitting routine are large (r.m.s.  residual $\approx 5.3\
\sigma_{\rm background}$) but do not appear to be systematic. 

Having just obtained reliable global scale parameters for our sample
galaxies, we are now ready to quantify the disc truncations occurring in
these galaxies. 

\section{Truncated discs}
\label{results.sect}

\subsection{Approach}
\label{radlum.sect}

In the analysis of edge-on galaxies, the inner disc region closest to
the plane often needs to be avoided because of the presence of either a
prominent dust lane, or a patchy dust distribution with its highest
density towards the galactic plane.  In many cases, the dust component
extends all the way to the edge of the disc, thus making the luminosity
distribution near the galactic planes useless for our study. 

The exponential scale{\it height} of galactic discs is -- to first order
-- constant as a function of galactocentric distance, at least for
later-type disc galaxies (see, e.g., KS1--4, Kylafis \& Bahcall 1987,
Shaw \& Gilmore 1990, Barnaby \& Thronson 1992, de Grijs \& Peletier
1997).  Therefore, the radial luminosity distributions parallel to the
galactic planes show a similar functional behaviour as the luminosity
profiles {\it in} the plane in the absence of the dust component. 
Consequently, by extracting light profiles parallel to the galactic
planes, we will also be able to study the occurrence and properties of
radially truncated discs, provided that the S/N ratio allows us to
detect such a truncation. 

\begin{figure}
\psfig{figure=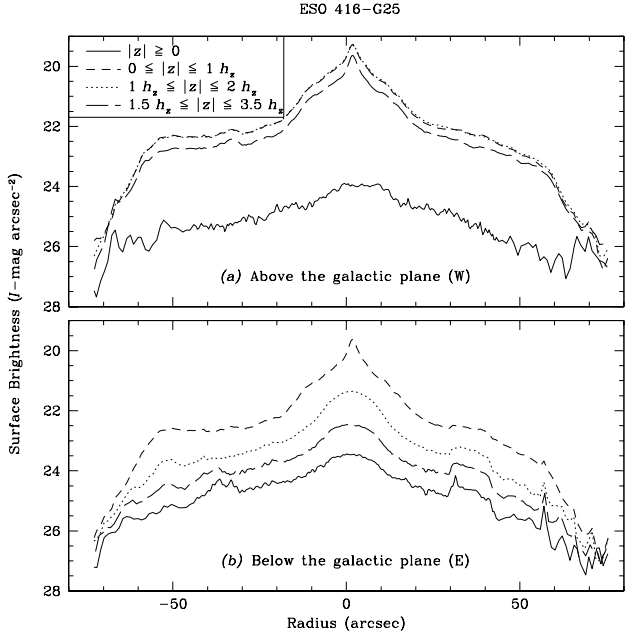,width=8.8cm}
\caption[]{\label{zrange.fig}Vertically averaged radial surface
brightness profiles of ESO 416-G25, taken at various {\it z}-heights
from the galactic plane. To retain an approximately constant S/N ratio,
a semi-logarithmic intensity-weighted radial binning algorithm has been
applied to the individual profiles. An indication of the observational
uncertainites is given by the background noise dominating the individual
profiles at large radii.}
\end{figure}

To find the optimum balance between avoiding contamination by the
in-plane dust component on the one hand, and retaining a sufficiently
high S/N ratio at large galactocentric distances on the other, numerous
experiments were performed.  In Fig.  \ref{zrange.fig} we show ESO
416-G25 as an example, where we compare several radial profiles obtained
on either side of the galactic plane.  The solid lines in both panels
represent the total profiles, vertically averaged over the entire half
of the galactic disc (effectively for $|z| \le 8 h_z$).  They are
obviously significantly affected by patchy dust and/or low S/N regions
throughout the disc and are therefore discarded from further use.  From
this figure (and similar figures for the other galaxies in our sample)
it follows that either the range $(1.0 \le |z| \le 2.0 h_z)$ or $(1.5
\le |z| \le 3.5 h_z)$ is to be preferred for our detailed analysis. 
Since the former region has in general a higher S/N ratio we conclude
that the most representative, unobscured light profiles suitable for a
further study of disc truncations are obtained by vertically collapsing
the surface brightness distribution between 1 and 2 $h_z$ {\it on the
least obscured side of the galactic plane}.  (Even though this galaxy
does not have a prominent dust lane nor a very large amount of patchy
extinction throughout its disc, Fig.  \ref{zrange.fig} clearly shows the
rationale behind choosing the least obscured side of the galactic disc
[top panel].)

In the radial direction, we apply a semi-logarithmic binning algorithm,
in order to retain an approximately constant overall S/N ratio (cf.  de
Grijs et al.  1997, de Grijs 1998), where the binning at the outermost
disc radii never exceeds 2 resolution elements. 

In the following sections, we will use profiles thus obtained for the
detailed analysis of the truncated discs in our pilot sample.

A potential problem of this method is that the S/N ratios in the outer
disc regions are often significantly lower at some (vertical) distance
from the galactic planes compared to those in the planes.  From Fig. 
\ref{zrange.fig} (and similar figures for the other sample galaxies) we
conclude, however, that although the radial extent of the cut-off region
appears to be a (weak) function of the height from the planes, the
actual radii at which the luminosity profiles disappear asymptotically
into the background noise converge to the same value of $R_{\rm max}$,
within the observational uncertainties. 

Secondly, some evidence exists that galactic discs thicken with
increasing galactocentric distance (e.g., KS1, Kent, Dame \& Fazio 1991,
Barnaby \& Thronson 1992, de Grijs \& van der Kruit 1996, de Grijs \&
Peletier 1997).  The signature of such a thickening of the discs on
light profiles extracted parallel to the galaxies' major axes is either
a flattening of the radial surface brightness profiles (if the
thickening occurs gradually; e.g., Kent et al.  1991, Barnaby \&
Thronson 1992, de Grijs \& Peletier 1997) or a locally enhanced surface
brightness level at these large galactocentric distances (if only the
outermost profiles are affected; e.g., KS1, de Grijs \& van der Kruit
1996).  However, de Grijs \& Peletier (1997) have shown that the effects
of disc thickening are largest for the earliest-type spiral galaxies and
almost zero for the later types, {\it including those examined in detail
in this paper.} Moreover, even though the discs of our sample galaxies
may have larger scaleheights with increasing radii, the low S/N ratios
and deviations from exponentially decreasing light distributions due to,
e.g., spiral arms will likely hide such observational signatures. 
Finally, the possible thickening of galactic discs will {\it not} affect
the determination of the actual truncation {\it radius} $R_{\rm max}$,
since the {\it radial} surface brightness distribution remains
unaffected. 

Although line-of-sight projection affects the observed functional form
of the radial luminosity distribution, we will use a version of Eq.  (2)
with an exponentially decreasing functionality in the truncation region,
but not including line-of-sight projection effects to fit {\it the
truncation regions} of our sample galaxies.  We chose to do so, because
(i) Eq.  (2) is a mathematically convenient function, and (ii) the
actual profile shapes in the truncation region are erratic due to low
S/N ratios and therefore the assumption of any more complicated
functionality than an exponentially decreasing luminosity density cannot
be taken seriously.  We will perform the actual fits to our
sky-subtracted images {\it in the linear regime} (i.e., in luminosity
instead of surface brightness space), to avoid undefined surface
brightness values due to ``negative'' noise peaks. 

\subsection{Artificial truncations?}

In interpreting the steep luminosity decline as truly representative of
a decrease in either the light or density distributions of a galactic
disc, one has to make sure that the observed cut-off is not an artifact
due to inaccurate sky subtraction.  De Vaucouleurs (1948), de
Vaucouleurs \& Capaccioli (1979), and van Dokkum et al.  (1994) have
shown that inaccurate sky subtraction (i.e., oversubtraction) causes a
false cut-off in the luminosity distribution of a galactic disc.  This
can easily be checked, because the artificial cut-off would not only be
present in a major axis cut, but also in cuts taken in other directions. 

In all cases, the background emission in the field of view of our sample
galaxies could be well represented by a plane, of which the slope was
determined by the flux in regions sufficiently far away from the
galaxies in order not to be affected by residual galactic light.  For
most of our observations, these planes could be closely approximated by
constant flux levels across the CCD field.  The remaining uncertainties
in the background are largely due to poisson noise. 

\begin{figure}
\psfig{figure=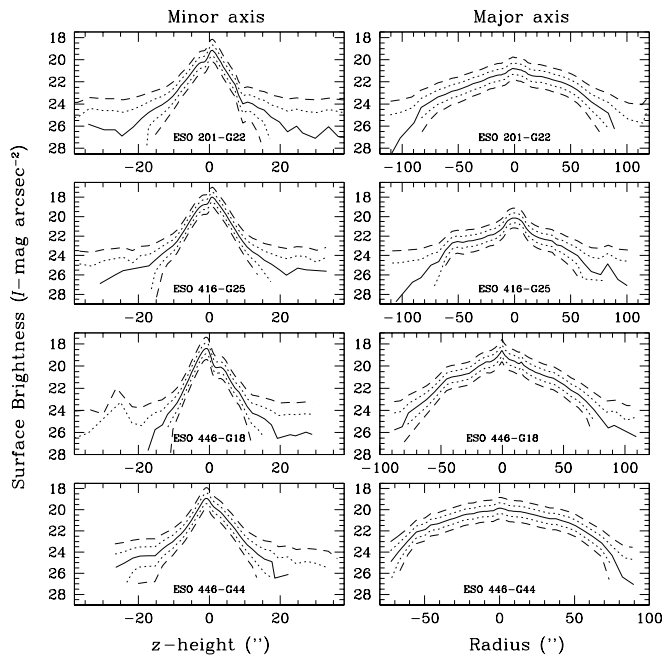,width=9cm}
\caption[]{\label{bgcheck.fig}Minor-axis (left-hand panels) and
major-axis (right-hand panels) {\it I}-band luminosity profiles of our
sample galaxies.  From top to bottom, the lines in each panel represent
the minor axis profiles $- 2 \sigma$ (dashed), $- 1 \sigma$ (dotted),
the profiles after subtraction of our best estimates for the sky
background, $+ 1 \sigma$ (dotted lines), and $+ 2 \sigma$ (dashed
lines), where $\sigma$ corresponds to the sky noise in the regions that
were used to determine the sky background levels.  For reasons of
clarity, the dashed and dotted profiles are displaced by, respectively,
$\pm 1.0$ and $\pm 0.5$ mag from the solid lines.}
\end{figure}

Fig.  \ref{bgcheck.fig} illustrates the quality of the background
subtraction in the {\it I} band, where the background contribution is
greater than in our other optical passbands.  The left-hand panels show
the minor-axis (vertical) surface brightness profiles of all sample
galaxies (solid lines), radially averaged over $\sim 20''$ in order to
be able to reach similar or fainter light levels as for the profiles
along the major axes, shown in the right-hand panels.  We determined the
sky noise, $\sigma$, in the regions used for the background subtraction
and created new images by subtracting (background $- 2 \sigma$),
(background $- 1 \sigma$), (background $+ 1 \sigma$), and (background $+
2 \sigma$), where ``background'' represents our best estimate of the sky
background in the individual images. The under and oversubtracted
profiles are shown offset from the solid profiles for reasons of
clarity. 

The effects of oversubtraction can clearly be seen in the minor-axis
surface brightness profiles: they show artificial cut-offs and the
negative background values result in undefined surface brightnesses at
these {\it z} heights, above or below $\approx 15''$ (i.e.  the profiles
could not be plotted beyond $\approx 15''$ due to undefined surface
brightness values resulting from oversubtraction of the sky background). 
Although the effects of oversubtraction on the major-axis profiles
(right-hand panels; vertically averaged between $-1.0$ and $1.0 h_z$)
show similar false cut-offs as for the minor-axis profiles, it appears
that most of the features seen in the light profiles represented by the
solid lines are real, since they are also observed in the {\it
under$\,$}subtracted light profiles.  Moreover, a qualitative comparison
of both the major and the minor-axis profiles (solid lines) shows that
the apparent truncations in or steepening of the major-axis light
profiles do not correspond to similar cut-offs at the same surface
brightness levels in the minor-axis profiles in any of our sample
galaxies.  We thus conclude that these features are not due to
inaccurate sky subtractions, but represent real deviations from the
radial exponential light profiles. 

Alternatively, the detection of radially truncated discs can be
artificially enhanced if the discs are strongly warped.  In fact, it
appears that for a number of our sample galaxies the locus of maximum
intensity at large radii may slightly deviate from the main galactic
plane direction (de Grijs 1997).  However, this effect is negligible for
the determination of their radial truncations, because the deviations
are {\it almost insignificant} and we average the radial luminosity
profiles over a sufficiently large vertical range to avoid such
problems. 

In addition, effects due to the galaxies' positions near the CCD edge or
to scattered light from foreground stars are potentially serious.  For
our four sample galaxies, the former are non-existent.  As one can see
in Fig.  \ref{contours.fig}, both ESO 201-G22 and ESO 446-G18 suffer
from the superposition of foreground stars, but only in the case of ESO
446-G18 this precludes us from determining its maximum disc radius on
the northern end; the superposed foreground star near the eastern edge
of ESO 201-G22 is located sufficiently far away from the truncation
region, {\it and} is found on the most obscured side of the galactic
plane. 

\subsection{Properties of truncated discs}
\label{properties.sect}

\subsubsection{Where do the truncations occur?}

\begin{figure}
\psfig{figure=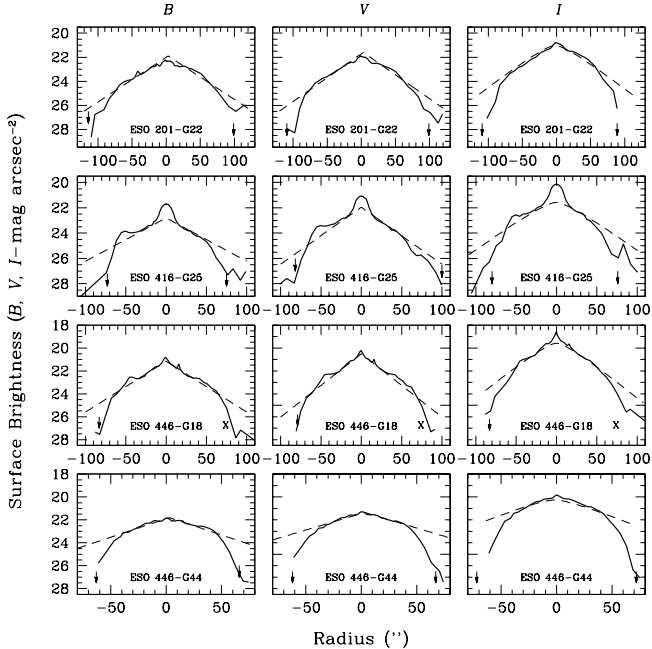,width=9cm}
\caption[]{\label{cutoffs.fig}Radial surface brightness profiles of our
sample galaxies taken parallel to their major axes (Sect. 
\ref{radlum.sect}).  Overplotted are the model exponential profiles
(dashed lines), based on our full surface brightness modelling.  The
arrows indicate the measured truncation radii; for ESO 446-G18 the
crosses indicate the side where we cannot determine the maximum radius
due to the presence of foreground stars.}
\end{figure}

In Fig.  \ref{cutoffs.fig} we show the radial surface brightness
profiles parallel to the major axis of our sample galaxies in all
passbands and out to either the edge of the CCD frames or to those radii
where the background noise dominates.  The dashed lines indicate
infinite exponential model discs, using the scale parameters obtained in
Sect.  \ref{2D.sect}.  It is immediately clear that the observed surface
brightness profiles are significantly fainter in the outer regions than
the model discs for all galaxies in our sample and for all passbands. 
(Note that since these profiles are only meant to guide the eye, we have
not computed the full line-of-sight integrated model profiles, although
such models were, in fact, used to obtain the actual scale parameters.)

The values for both $R_{\max}$ and $\delta$, resulting from the fitting
of our modified version of Eq.  (2) to the observed profiles using a
standard linear least-squares fitting technique, are listed in Table
\ref{Rmax.tab}.  Although the formal measurement errors in $R_{\rm max}$
are $\le 2''$, the relatively large uncertainties associated with
$R_{\rm max}$ are due to the fact that S/N $< 1$ at the truncation
radius (by definition), and to the difficulty in the unambiguous
determination of the start of the truncation region.  In fact, the
uncertainties in the truncation lengths are {\it predominantly}
determined by the nature of $R_{\rm max}$ as lower limit. 

However, this method will at least produce objective estimates of
$R_{\rm max}$, as opposed to visually extrapolating the observed surface
brightness profiles to those radii where they would (supposedly)
disappear asymptotically into the noise, as has been done previously by
other workers in this field. 

For a comparison with previously published results for other galaxies,
we have also included the estimated truncation radii in units of the
galaxies' {\it I}-band scalelengths.  We chose to use the {\it I}-band
scalelengths to determine the $R_{\rm max} / h_R$ ratios, because these
represent our longest-wavelength observations, which most likely best
approximate the dominant stellar luminosity (and presumably mass)
distributions (de Grijs et al.  1997, de Grijs 1998).  The corresponding
errors reflect the uncertainties in the determinations of both the
scalelengths and the truncation radii. 

Van der Kruit \& Searle (KS3) determined, for their small sample of
large edge-ons, that the mean truncation radius $R_{\rm max} \simeq (4.2
\pm 0.6) h_R$ (cf.  Bottema 1995 for NGC 4013: $R_{\rm max} \simeq 4.1
h_R$).  Barteldrees \& Dettmar (1994) found, for a sample of 27 edge-on
galaxies, a mean truncation radius of $\simeq (3.7 \pm 1.0) h_R$. 
However, this result is based on a different definition of $R_{\rm
max}$: they interpreted the truncation radius as the galactocentric
distance at which the observed projected radial profiles start to
deviate significantly from the model exponential profiles.  If we keep
in mind that the truncation occurs over a finite region, then the
discrepancy between these determinations can be understood.  A direct
comparison is therefore impossible. 

Recently, Pohlen et al.  (2000a) largely reanalysed Barteldrees \&
Dettmar's (1994) sample, assuming infinitely sharply truncated galactic
discs following KS3.  For their sample of 31 nearby edge-on spiral
galaxies, they found $R_{\rm max}/h_R = 2.9 \pm 0.7$, significantly
lower than the ratio found by KS3.  This may reflect a selection bias
towards large galaxies and/or small-number statistics in the KS3 sample. 

While the discs of ESO 201-G22, ESO 416-G25, and ESO 446-G18 are
truncated at comparable radii as found by KS and Bottema (1995), expressed
in units of their disc scalelengths, ESO 446-G44 is clearly truncated at
much smaller radii.  This is the sample galaxy with the greatest
scalelength and the only one without a bulge. ESO 446-G44, as well as
ESO 416-G25, exhibit truncated discs well within the range found by
Pohlen et al. (2000a).

\begin{figure}
\psfig{figure=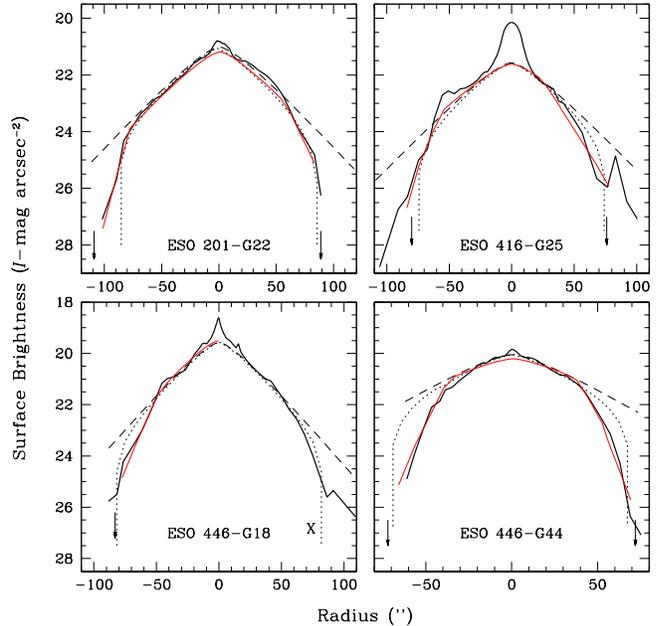,width=9cm}
\caption[]{\label{Iband.fig}{\it I}-band radial surface brightness
profiles of our sample galaxies taken parallel to their major axes, as
in Fig.  \ref{cutoffs.fig}.  Overplotted are the model exponential
profiles (dashed lines), based on our full surface brightness modelling. 
The arrows indicate the measured truncation radii; for ESO 446-G18 the
cross indicates the side where we cannot determine the maximum radius
due to the presence of foreground stars.  The dotted lines correspond to
{\it infinitely sharply truncated} models, used for a comparison with
previous work (Sect.  \ref{asymm.sect}); the thin solid lines are our
model fits using a slightly modified Casertano model (Sect. 
\ref{3Dmodel.sect}).}
\end{figure}

{
\begin{table*}

\caption[]{\label{Rmax.tab}{\bf Disc truncations in our sample galaxies}\\ 
Columns: (1) Galaxy name; (2) Passband observed in; (3) Side of the
galactic centre; (4) and (5) Cut-off radius (lower limit) and
observational error (arcsec); (6) and (7) Cut-off radius (in units of
the {\it I-}band scalelength) and observational error; (8) Truncation
length (arcsec; typical observational uncertainties are of order $\ge
10''$); (9) and (10) Scalelength in the truncation region and
observational error, in arcsec; (11) and (12) Scalelength in the
truncation region and observational error, in kpc (based on the
heliocentric velocities obtained by Mathewson, Ford \& Buchhorn 1992
[see de Grijs 1998])}

\begin{center}
\tabcolsep=1mm

\begin{tabular}{cccrrccccrcccc}
\hline
Galaxy & Band & Side & $R_{\rm max}$ & $\pm$ & & $R_{\rm max}$ &
$\pm$ & $\delta$ & $h_{R,\delta}$ & $\pm$ & & $h_{R,\delta}$ & $\pm$ \\
\cline{4-5}\cline{7-8}\cline{10-11}\cline{13-14}
\noalign{\vspace{2pt}}
(ESO-LV) & & & \multicolumn{2}{c}{$('')$} & & \multicolumn{2}{c}{$(h_{R,I})$} &
$('')$ & \multicolumn{2}{c}{$('')$} & & \multicolumn{2}{c}{($h^{-1}$ kpc)} \\
(1) & (2) & (3) & (4) & (5) & & (6) & (7) & (8) & (9) & (10) & & (11) & (12) \\
201-G22 & {\it B} & E & 114 & 10 & & 4.9 & 0.4 & 44 & 14.2 & 1.0 & & 3.0 & 0.2 \\
        &         & W &  99 &  4 & & 4.2 & 0.2 & 50 & 16.8 & 0.9 & & 3.6 & 0.2 \\
        & {\it V} & E & 109 &  3 & & 4.7 & 0.1 & 26 & 16.4 & 0.3 & & 3.5 & 0.1 \\
        &         & W &  99 & 10 & & 4.2 & 0.4 & 50 & 15.7 & 0.4 & & 3.4 & 0.1 \\
        & {\it I} & E & 109 & 10 & & 4.7 & 0.4 & 28 & 17.6 & 0.3 & & 3.8 & 0.1 \\
        &         & W &  89 &  8 & & 3.8 & 0.3 & 31 & 12.9 & 0.2 & & 2.8 & 0.1 \\
416-G25 & {\it B} & N &  73 &  5 & & 3.3 & 0.2 & 13 &  6.2 & 1.6 & & 1.7 & 0.4 \\
        &         & S &  75 & 10 & & 3.3 & 0.5 & 30 & 10.6 & 1.4 & & 2.9 & 0.4 \\
        & {\it V} & N &  82 & 10 & & 3.7 & 0.5 & 24 &  7.2 & 1.0 & & 2.0 & 0.3 \\
        &         & S & 100 & 15 & & 4.5 & 0.7 & 60 & 14.9 & 0.3 & & 4.1 & 0.1 \\
        & {\it I} & N &  80 & 15 & & 3.6 & 0.7 & 28 &  9.2 & 1.7 & & 2.6 & 0.4 \\
        &         & S &  76 & 10 & & 3.4 & 0.5 & 53 & 14.3 & 1.3 & & 4.0 & 0.4 \\
446-G18 & {\it B} & S &  83 &  8 & & 4.6 & 0.4 & 35 & 11.6 & 0.6 & & 2.8 & 0.1 \\
        & {\it V} & S &  79 &  8 & & 4.4 & 0.4 & 31 & 10.3 & 0.3 & & 2.5 & 0.1 \\
        & {\it I} & S &  83 &  8 & & 4.6 & 0.4 & 35 &  9.8 & 1.0 & & 2.4 & 0.2 \\
446-G44 & {\it B} & E &  66 &  8 & & 2.3 & 0.3 & 25 & 10.2 & 0.2 & & 1.6 & 0.1 \\
        &         & W &  63 &  8 & & 2.1 & 0.3 & 22 &  7.6 & 0.8 & & 1.2 & 0.1 \\
        & {\it V} & E &  67 &  8 & & 2.3 & 0.3 & 27 &  8.6 & 1.3 & & 1.3 & 0.3 \\
        &         & W &  62 &  8 & & 2.1 & 0.3 & 22 &  7.6 & 0.8 & & 1.2 & 0.2 \\
        & {\it I} & E &  72 &  7 & & 2.4 & 0.3 & 33 &  9.8 & 0.3 & & 1.5 & 0.1 \\
        &         & W &  72 &  8 & & 2.4 & 0.3 & 32 &  8.6 & 0.3 & & 1.3 & 0.1 \\
\hline
\end{tabular}
\end{center}
\end{table*}
}

\subsubsection{Asymmetry and sharpness}
\label{asymm.sect}

Disc truncations do not necessarily occur at the same galactocentric
distances or with the same abruptness on either side (e.g., KS1, Jensen
\& Thuan 1982, N\"aslund \& J\"ors\"ater 1997, Abe et al.  1999, Fry et
al.  1999).  In most cases, however, the truncation radii on either side
of the galactic disc occur within $\sim 10-15$\% of each other.  The
profiles in Figs.  \ref{cutoffs.fig} and \ref{Iband.fig}, and our
estimates for $R_{\rm max}$ in Table \ref{Rmax.tab} show that in
general, the discs of our sample galaxies are truncated at similar
radii, within their observational uncertainties, with the possible
exception of ESO 201-G22. 

Table \ref{Rmax.tab} also contains our best estimates for the
exponential scalelength in the truncation region, $h_{R,\delta}$.  To
measure this scalelength, we defined the inner fitting radius as the
radius where the radial surface brightness profile starts to deviate
significantly from the model radial exponential light distribution (cf. 
Barteldrees \& Dettmar 1994); the uncertainty in this radius is
generally $\le 15$\%, depending on the particular galaxy considered.  As
outer fitting boundary we used $R_{\rm max}$.  The errors associated
with $h_{R,\delta}$ are observational uncertainties, obtained from the
comparison of several similar fits in which we adjusted the inner
boundary of the radial fitting range by 10--20\%.  The uncertainties in
$R_{\rm max}$ are included in the observational errors given; the formal
errors were $\le 1$\%.  The effects on disc truncations and scale
parameters of an inclination $i \neq 90^\circ$ are negligible for our
galaxy sample of $i \ge 87^\circ$ inclined galaxies, as was convincingly
shown by Barteldrees \& Dettmar (1994), in particular in view of the
systematic and observational uncertainties involved.  The combination of
$h_{R,\delta}$ and $\delta$ gives us an indication of the asymmetry and
sharpness of the actual disc truncations. 

Considering the relatively large systematic and observational errors
that cannot be avoided at this point, we cannot claim that we detect any
systematic asymmetries, perhaps with the exception of ESO 416-G25.  The
northern edge of the disc of ESO 416-G25 is very sharply truncated
compared to its southern edge.  Upon close examination of the actual CCD
images, we believe that this may be explained by the fact that we likely
observe the outer stellar envelope of a spiral arm, whereas on the
southern side we are looking into the inside of a spiral arm.  Note,
however, that also at the southern edge of the disc a clear truncation
signature is observed (Fig.  \ref{cutoffs.fig}). 

The last two columns of Table \ref{Rmax.tab} show that in none of our
sample galaxies the radial scalelength in the truncation region
decreases to values of order or less than 1 kpc.  Using $H_0 = 50$ km
s$^{-1}$ Mpc$^{-1}$, or other values closer to the current best
estimate, will further increase the truncation scalelengths measured in
our galaxies.  Although two of our sample galaxies may not be exactly
edge-on, their inclinations are sufficiently close to 90$^\circ$ so as
not to increase the measurements of $h_{R,\delta}$ by more than their
observational uncertainties (cf.  Barteldrees \& Dettmar 1994).  We are
therefore forced to conclude that, although our discs are clearly
truncated, the truncation occurs over a larger region and not as
abruptly as found in previous studies. 

As a comparison between our approach and that used in previous work, in
Fig.  \ref{Iband.fig} we present enlarged versions of the {\it I}-band
profiles of Fig.  \ref{cutoffs.fig}, in which we show both our best-fit
radial profiles using a slightly modified Casertano model (Sect. 
\ref{3Dmodel.sect}; thin solid lines) and the corresponding profiles for
a symmetric, projected, sharply truncated exponential disc, with the
disc truncations occurring at $R_{\rm max}$ (dotted lines).  These
profiles were obtained by applying the same method used to extract the
observed profiles used for Fig.  \ref{cutoffs.fig} to model images of
our sample galaxies.  From a comparison between these dotted lines and
the actual, observed profiles, it is clear that the discs of our sample
galaxies are generally {\it not infinitely sharply truncated}, but show
a more gradual decrease of their radial luminosity density. 

\begin{figure}
\psfig{figure=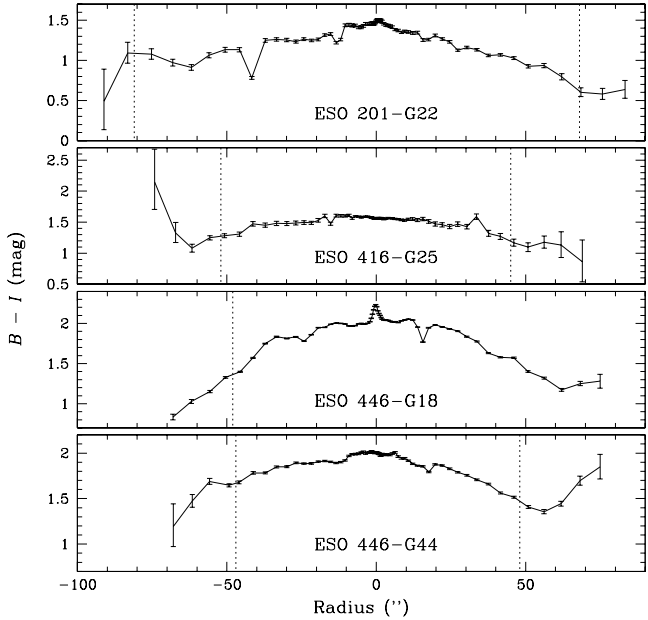,width=9cm}
\caption[]{\label{truncolors.fig}Radial $(B-I)$ colour profiles of our
sample galaxies parallel to their major axes, constructed from the
profiles shown in Fig. \ref{cutoffs.fig}. The dashed lines indicate the
approximate start of the truncation regions. All data points with
colour uncertainties $\sigma_{(B-I)} \le 0.4$ mag have been included.}
\end{figure}

Finally, from a galaxy-by-galaxy inspection of the values for $R_{\rm
max}$, no apparent trend with wavelength can be discerned, within the
observational errors.  However, a detailed comparison of the radial $(B
- I)$ colour profiles, our longest colour baseline, reveals that the
disc colour tends to get bluer in the truncation region compared to the
colours in the main disc (Fig.  \ref{truncolors.fig}).  A similar result
was obtained by Sasaki (1987) for the truncation region of NGC 5907. 
Although this may be indicative of more recent star formation at the
edge of the discs (e.g., Larson 1976, Seiden et al.  1984), the opposite
behaviour is exhibited both on the northern side of the disc of ESO
416-G25, and on the western side of the disc of ESO 446-G44, where the
colour reddens in the truncation region.  This is confirmed by the fact
that at these edges, these discs appear to be more sharply truncated in
the {\it B} band compared to the {\it I-}band observations, which may be
the signature of a spiral arm. 

\section{Dynamical consequences of disc truncations: Outlook}
\label{dynamics.sect}

Although the fact that many spiral discs seem to have truncated stellar
discs is an interesting observation in the context of galactic
structure, the physical implications of a similar truncation in the mass
distribution of galactic discs has far-reaching consequences making the
study of disc truncations fundamental to our understanding of galactic
disc maintenance and evolution.  This is therefore the main science
driver of our attempts to define a unique and objective method to
measure disc truncations. 

\subsection{Edge smearing and disc asymmetries}

The persistence of sharp disc cut-offs places a strong upper limit on
the stellar velocity dispersion at the disc edge (KS1).  Adopting a
rotational velocity of 250 km s$^{-1}$ at 20 kpc for NGC 4565, the
radial stellar velocity dispersion, $\langle v^2_R \rangle^{1/2}$, must
be $\le 10$ km s$^{-1}$ so that random motions do not wash out the sharp
cut-off within one revolution time (Jensen \& Thuan 1982; see also KS1,
May \& James 1984), or the sharp cut-off must be a transient feature
(e.g., Sasaki 1987).  This low upper limit for $\langle v^2_R
\rangle^{1/2}$ is close to the minimum value of $\sim 2$ km s$^{-1}$
needed to satisfy Toomre's (1964) criterion of local stability for disc
galaxies, and thus for star formation.

Alternatively, in the case of a disc formation scenario in which the
disc grows from the inside outward (e.g., Larson 1976, Gunn 1982, Seiden
1983, Seiden et al.  1984) a sharp edge can be maintained if this
outward growth is sufficiently rapid, so that the random motion of the
stars does not smear out the edge.  Note, however, that the disc
truncations in our sample galaxies are not as sharp as those found by
KS1--4, among others, which will relax these requirements. 

The situation becomes more complicated if the galactic disc is lopsided
or if the truncations occur at different radii.  Following the epicyclic
description of Baldwin, Lynden-Bell \& Sancisi (1980), van der Kruit
(1988) estimates a smearing time of $1.7 \times 10^{10}$ yr for the
Galactic disc, and he concludes that a variation in the truncation radii
of order 10\% may just survive a Hubble time.  With the possible
exception of ESO 416-G25, our sample galaxies appear to comfortably meet
this requirement. 

\subsection{Rotation curves as diagnostic tools}

Casertano (1983) has shown that a truncated stellar disc leaves a
signature on the rotation curve in the form of a region of slowly
varying velocity followed by a steep decline just outside the truncation
radius (see also Hunter, Ball \& Gottesman 1984).  The amount of this
decrease is a measure of the disc mass.  The effect of a truncation is a
{\it flattening} of the rotation curve inside the truncation itself,
from some radius $R_0$ to $R_{\rm max}$, and a steep decrease of the
velocity outside. 

The well-known warped edge-on galaxy NGC 4013, for which Bottema (1995)
suspected a sudden decrease in the mass density corresponding to the
truncation radius, has indeed been shown to exhibit a sudden drop in the
rotational velocity of about 20 km s$^{-1}$ just at the optical edge
(Bottema, Shostak \& van der Kruit 1987, Bottema 1995, 1996).  This drop
can be understood if one realises that near the edge of the galactic
disc the mass distribution will be irregular: there is no smooth,
circular end to the disc, but it likely ends in spiral arms.  Bottema
(1996) argues that therefore gas moving in the potential of such patches
of stellar matter will not be in precise circular motion and hence the
radial velocity along the line of sight is somewhat lower than the true
rotation. 

Finally, Bahcall (1983) showed that, for Sb or Sc galaxies like NGC 891
or the Galaxy, the feature in the rotation curve due to the truncated
stellar disc is observable only if $R_{\rm max} \le 4 h_R$ (smaller for
galaxies with more prominent bulges), if the truncation length is small
compared to $h_R$, and if the halo mass inside $R_{\rm max}$ is smaller
than the disc mass (Casertano 1983). 

Unfortunately, the currently available velocity information for the four
galaxies in our pilot sample does not allow us to confirm the presence
of sharp truncations in the disc mass based on the shape of the rotation
curves: only for ESO 446-G18 and ESO 446-G44 rotation curves have been
published, for the H$\alpha$ emission (Mathewson et al.  1992) and the
H{\sc i} component (Persic \& Salucci 1995, based on the raw Mathewson
et al.  1992 data), but these rotation curves do not or just barely
reach those radii where we expect to be able to see a truncation
signature.

\section{Summary and Conclusions}

In this paper we have presented the first results of a systematic
analysis of galactic disc structure in general and of radially truncated
exponential discs in particular for a pilot sample of four ``normal''
disc-dominated edge-on spiral galaxies.  We have carefully considered
the importance of (residual) dust, deviations from 90$^\circ$
inclinations, and spiral arms, and concluded that these effects do not
affect our results significantly.  We have also shown that the truncated
discs in our sample galaxies are not caused artificially by inaccurate
sky subtraction, but are real deviations from the radial exponential
light profiles. 

An independent approach to obtain the statistics of truncated galactic
discs, using a sample of galaxies selected in a uniform way, is needed
in order to better understand their overall properties and physical
implications.  If the truncations seen in the stellar light are also
present in the mass distribution, they would have important dynamical
consequences at the disc's outer edges.  We have shown that the
truncated luminosity distributions of our pilot sample galaxies, if also
present in the mass distributions, comfortably meet the requirements for
longevity. 

The truncation radii, expressed in units of $h_R$, for the discs of ESO
201-G22, ESO 416-G25, and ESO 446-G18 are comparable to those found by
KS1--4 and Bottema (1995), while ESO 446-G44 is truncated at much
smaller radii.  In fact, the truncations of the discs of ESO 416-G25 and
ESO 446-G44 are within the range found by Pohlen et al.  (2000a) for
their sample of 31 nearby edge-on spiral galaxies.  In general, the
discs of our sample galaxies are truncated at similar radii on either
side of their centres, within the observational uncertainties, with the
exception of ESO 201-G22. 

With the possible exception of the disc of ESO 416-G25, it appears that
our sample galaxies are fairly symmetric, in terms of both the sharpness
of their disc truncations and the truncation length, although the
truncations occur over a larger region and not as abruptly as found in
previous studies.  The northern edge of the disc of ESO 416-G25 is very
sharply truncated compared to its southern edge.  We believe that this
may be explained by the fact that we likely observe the outer stellar
envelope of a spiral arm, whereas on the southern side we are looking
into the inside of a spiral arm. 

\section*{Acknowledgments} We thank Piet van der Kruit for stimulating
discussions and acknowledge useful suggestions by the referee, M. 
Pohlen.  This work is partially based on the undergraduate senior thesis
of KHW at the University of Virginia.  RdeG acknowledges partial funding
from NASA grants NAG 5-3428 and NAG 5-6403 and hospitality at the
University of Groningen.  This research has made use of NASA's
Astrophysics Data System Abstract Service and of the NASA/IPAC
Extragalactic Database (NED).

\end{document}